\documentclass[aps,prb,twocolumn,showpacs,superscriptaddress]{revtex4}

\usepackage{graphicx}
\usepackage{amsmath}
\usepackage{amssymb}

\begin{document}

\title{Pseudogap-induced asymmetric tunneling in cuprate superconductors}

\author{L\"ulin Kuang}

\affiliation{Department of Physics, Beijing Normal University, Beijing 100875, China}

\author{Huaisong Zhao}

\affiliation{College of Physics, Qingdao University, Qingdao 266071, China}

\author{Shiping Feng}

\affiliation{Department of Physics, Beijing Normal University, Beijing 100875, China}

\begin{abstract}
Within the framework of the kinetic energy driven superconducting mechanism, the doping and temperature dependence of the asymmetric tunneling in cuprate superconductors is studied by considering the interplay between the superconducting gap and normal-state pseudogap. It is shown that the asymmetry of the tunneling spectrum in the underdoped regime weakens with increasing doping, and then the symmetric tunneling spectrum recovers in the heavily overdoped regime. The theory also shows that the asymmetric tunneling is a natural consequence due to the presence of the normal-state pseudogap.
\end{abstract}


\maketitle

Cuprate superconductors are complex materials that exhibit a variety of phases determined not only by temperature but also by charge carrier doping \cite{Batlogg94,Timusk99,Hufner08}. The pairing of electrons in the conventional superconductors \cite{Schrieffer64} occurs at the superconducting (SC) transition temperature
$T_{\rm c}$, creating an energy gap in the electron excitation spectrum that serves as the SC order parameter. However, in cuprate superconductors, the normal-state pseudogap exists between $T_{\rm c}$ and the temperature $T^{*}$, with $T^{*}$ is called as the normal-state pseudogap crossover temperature \cite{Batlogg94,Timusk99,Hufner08}. Although $T_{\rm c}$ takes a domelike shape with the underdoped and overdoped regimes on each side of the optimal doping, where $T_{\rm c}$ researches its maximum \cite{Tallon95}, $T^{*}$ is much larger than $T_{\rm c}$ in the underdoped regime, then it monotonically decreases with increasing doping, and seems to merge with the $T_{\rm c}$ in the overdoped regime, eventually disappearing together with superconductivity at the end of the SC dome \cite{Batlogg94,Timusk99,Hufner08}. After intensive investigations over more than two decades, it has become clear that many of the unusual physical properties in cuprate superconductors can be attributed to the emergence of the normal-state pseudogap \cite{Batlogg94,Timusk99,Hufner08}.

The complexity in cuprate superconductors is reflected in the quasiparticle excitation spectra \cite{Damascelli03,Fujita12,Fischer07}. The scanning tunneling microscopy/spectrascopy (STM/STS) is a powerful tool to study the quasiparticle properties in cuprate superconductors \cite{Fujita12,Fischer07}, since its remarkable energy and spatial resolution makes it particularly well suited for cuprate superconductors, which are characterized by small energy and short length scales. More accurately, the STM/STS data are proportional to the local density of quasiparticle excitations, and the accounting of their distribution can provide important insight into the nature of cuprate superconductors. In the conventional superconductors, the most complete and convincing evidence for the electron-phonon SC mechanism came from the tunneling spectrum \cite{Schrieffer64,Giaever60}. During the last two decades, the tunneling study of cuprate superconductors has revealed many crucial results \cite{Fujita12,Fischer07,Renner98,Pan00,Hanaguri04,Kugler06}, where the main feature of the differential tunneling conductance spectrum is the quasiparticle excitation gap. Moreover, the presence of the excitations within the SC gap, linearly increasing with energy around ${\rm V}=0$, indicates that the SC gap has nodes, and therefore presumably d-wave symmetry. In particular, the most remarkable feature about the tunneling in cuprate superconductors is the fact that the tunneling conductivity between a metallic point and a cuprate superconductor is markedly asymmetric between positive and negative voltage biases \cite{Anderson06}.

A challenging issue for theory is to explain the asymmetric tunneling in cuprate superconductors. Recently, we \cite{Feng12} have discussed the interplay between the normal-state pseudogap state and superconductivity in cuprate superconductors within the framework of the kinetic energy driven SC mechanism \cite{Feng0306}, where both the charge carrier pairing state in the particle-particle channel and normal-state pseudogap state in the particle-hole channel arise from the same interaction that originates directly from the kinetic energy by exchanging spin excitations, then there is a coexistence of the SC gap and normal-state pseudogap in the whole SC dome. Furthermore, both the normal-state pseudogap and the SC gap are dominated by one energy scale, and they are the result of the strong electron correlation. Within this microscopic SC theory, some unusual properties of cuprate superconductors in the pseudogap phase have been studied \cite{Zhao12}, including the humplike anomaly of the specific-heat, the particle-hole asymmetry electronic state, and the unusual evolution of the Fermi arc length with doping and temperature, and the results are qualitatively consistent with the experimental results. In this paper, we study the doping and temperature dependence of the asymmetric tunneling in cuprate superconductors along with this line. by considering the interplay between the normal-state pseudogap state and superconductivity, we qualitatively reproduce some main features of the STM/STS measurements on cuprate superconductors in the whole doping range from the underdoped to heavily overdoped \cite{Fujita12,Fischer07,Renner98,Pan00,Hanaguri04,Kugler06}. In particular, we show that the remarkably asymmetric tunneling in cuprate superconductors is a natural consequence due to the presence of the normal-state pseudogap.

Although there are hundreds of cuprate SC compounds, they all share a layered structure which contains one or more copper-oxygen planes \cite{Damascelli03}. In this case, it has been argued strongly \cite{Anderson87} that the low-energy physics of these planes is described by the two-dimensional $t$-$J$ model acting on the Hilbert space with no doubly occupied sites, where the kinetic energy includes the nearest-neighbor (NN) and next NN hopping on a square lattice with the matrix elements denoted as $t$ and $t'$, respectively, while the antiferromagnetic (AF) Heisenberg term with the exchange coupling constant $J$ describes the AF coupling between localized spins. To incorporate the electron motion within the restricted Hilbert space without double electron occupancy, we \cite{Feng04} have developed the charge-spin separation (CSS) fermion-spin theory, where the constrained electron operators are decoupled as $C_{l\uparrow}= h^{\dagger}_{l\uparrow}S^{-}_{l}$ and $C_{l\downarrow}=h^{\dagger}_{l\downarrow}S^{+}_{l}$, with the spinful fermion operator $h_{l\sigma}=e^{-i\Phi_{l\sigma}}h_{l}$ that describes the charge degree of freedom of the electron together with some effects of spin configuration rearrangements due to the presence of the doped hole itself (charge carrier), while the spin operator $S_{l}$ keeps track of the spin degree of freedom of the electron, then the electron single occupancy local constraint is satisfied in analytical calculations. In this CSS fermion-spin representation, the $t$-$J$ model can be expressed explicitly as,
\begin{eqnarray}\label{cssham}
H&=&t\sum_{l\hat{\eta}}(h^{\dagger}_{l+\hat{\eta}\uparrow}h_{l\uparrow}S^{+}_{l}S^{-}_{l+\hat{\eta}}+h^{\dagger}_{l+\hat{\eta}\downarrow}h_{l\downarrow}S^{-}_{l}
S^{+}_{l+\hat{\eta}})\nonumber\\
&-&t'\sum_{l\hat{\tau}}(h^{\dagger}_{l+\hat{\tau}\uparrow}h_{l\uparrow}S^{+}_{l}S^{-}_{l+\hat{\tau}}+h^{\dagger}_{l+\hat{\tau}\downarrow}
h_{l\downarrow}S^{-}_{l}S^{+}_{l+\hat{\tau}})\nonumber\\
&-&\mu\sum_{l\sigma} h^{\dagger}_{l\sigma}h_{l\sigma}+J_{{\rm eff}}\sum_{l\hat{\eta}}{\bf S}_{l}\cdot {\bf S}_{l+\hat{\eta}},
\end{eqnarray}
where the summations $l\hat{\eta}$ and $l\hat{\tau}$ are carried over NN and next NN bonds, respectively, ${\bf S}_{l}=(S^{\rm x}_{l},S^{\rm y}_{l}, S^{\rm z}_{l})$ are spin operators, $S^{-}_{l}$ and $S^{+}_{l}$ are the spin-lowering and spin-raising operators for the spin $S=1/2$, respectively, $\mu$ is the chemical potential,
$J_{{\rm eff}}=(1-\delta)^{2}J$, and $\delta=\langle h^{\dagger}_{l\sigma}h_{l\sigma}\rangle=\langle h^{\dagger}_{l}h_{l}\rangle$ is the charge carrier doping concentration.

Superconductivity, the dissipationless flow of electrical current, is a striking manifestation of a subtle form of quantum rigidity on the  macroscopic scale, where a central question is how the SC-state forms? It is all agreed that the electron Cooper pairs are crucial for the form of the SC-state because these electron Cooper pairs behave as effective bosons, and can form something analogous to a Bose condensate that flows without resistance \cite{Schrieffer64,Tsuei00}. This follows a fact that although electrons repel each other because of the Coulomb interaction, at low energies there can be an effective attraction that originates by the exchange of bosons. In the conventional superconductors, these exchanged bosons are {\it phonons} that act like a bosonic {\it glue} to hold the electron pairs together, then these electron Cooper pairs condense into a coherent macroscopic quantum state that is insensitive to impurities and imperfections and hence conducts electricity without resistance \cite{Schrieffer64}. For cuprate superconductors, we \cite{Feng0306} have shown in terms of Eliashberg's strong coupling theory \cite{Eliashberg66} that in the doped regime without an AF long-range order the charge carriers are held together in pairs in the particle-particle channel by the effective interaction that originates directly from the kinetic energy of the $t$-$J$ model (\ref{cssham}) by the exchange of {\it spin excitations}, then the electron Cooper pairs originating from the charge carrier pairing state are due to the charge-spin recombination, and their condensation reveals the SC ground-state. In particular, this SC-state is controlled by both SC gap and quasiparticle coherence, which leads to that the maximal $T_{\rm c}$ occurs around the optimal doping, and then decreases in both underdoped and overdoped regimes. Furthermore, this same interaction also induces the normal-state pseudogap state in the particle-hole channel \cite{Feng12}. Since this normal-state pseudogap is closely related to the quasiparticle coherent weight, and therefore it suppresses the spectral weight. Following our previous discussions \cite{Feng12,Feng0306}, the full charge carrier diagonal and off-diagonal Green's functions of the $t$-$J$ model (\ref{cssham}) in the SC-state are evaluated as,
\begin{widetext}
\begin{subequations}\label{hole-Green-function}
\begin{eqnarray}
g({\bf k},\omega)&=&{1\over \omega-\xi_{\bf k}-\Sigma^{({\rm h})}_{1}({\bf k},\omega)-\bar{\Delta}^{2}_{\rm h}({\bf k})/[\omega+\xi_{\bf k}
+\Sigma^{({\rm h})}_{1}({\bf k}, -\omega)]},\label{hole-diagonal-Green's-function}\\
\Gamma^{\dagger}({\bf k},\omega)&=&-{\bar{\Delta}_{\rm h}({\bf k})\over [\omega-\xi_{\bf k}-\Sigma^{({\rm h})}_{1}({\bf k},\omega)][\omega+\xi_{\bf k}
+\Sigma^{({\rm h})}_{1}({\bf k},-\omega)]-\bar{\Delta}^{2}_{\rm h}({\bf k})},
\end{eqnarray}
\end{subequations}
\end{widetext}
where $\xi_{\bf k}=Zt\chi_{1}\gamma_{{\bf k}}-Zt'\chi_{2}\gamma_{{\bf k}}'-\mu$ is the mean-field (MF) charge carrier spectrum with the spin correlation functions
$\chi_{1}=\langle S^{+}_{l}S^{-}_{l+\hat{\eta}}\rangle$ and $\chi_{2}=\langle S_{l}^{+}S_{l+\hat{\tau}}^{-}\rangle$,
$\gamma_{{\bf k}}=(1/Z)\sum_{\hat{\eta}}e^{i{\bf k}\cdot\hat{\eta}}$, $\gamma_{{\bf k}}'=(1/Z)\sum_{\hat{\tau}}e^{i{\bf k}\cdot\hat{\tau}}$, $Z$ is the number of the NN or next NN sites on a square lattice, the effective charge carrier pair gap $\bar{\Delta}_{\rm h}({\bf k})$ is closely associated with the self-energy $\Sigma^{({\rm h})}_{2}({\bf k},\omega)$ in the particle-particle channel as $\bar{\Delta}_{\rm h}({\bf k})=\Sigma^{({\rm h})}_{2}({\bf k},\omega=0)$, and can be expressed explicitly as a d-wave form
$\bar{\Delta}_{\rm h}({\bf k})=\bar{\Delta}_{\rm h}\gamma^{({\rm d})}_{{\bf k}}$ with $\gamma^{({\rm d})}_{{\bf k}}=({\rm cos} k_{x}-{\rm cos}k_{y})/2$, while the self-energy $\Sigma^{({\rm h})}_{1}({\bf k},\omega)$ in the particle-hole channel renormalizes the MF charge carrier spectrum, and can be rewritten approximately as
$\Sigma^{({\rm }h)}_{1}({\bf k},\omega)\approx {[2\bar{\Delta}_{\rm pg}({\bf k})]^{2}/[\omega+M_{\bf k}}]$, where $M_{\bf k}$ is the energy spectrum of
$\Sigma^{({\rm h})}_{1}({\bf k},\omega)$, and $\bar{\Delta}_{\rm pg}({\bf k})$ is the effective normal-state pseudogap. With these above definitions, the Green's functions in Eq. (\ref{hole-Green-function}) are obtained explicitly as \cite{Feng12},
\begin{subequations}\label{hole-Green-function-1}
\begin{eqnarray}
g(\bf k,\omega)&=&\sum_{\nu=1,2}\left ( {U^{2}_{\nu{\rm h}{\bf k}}\over\omega-E_{\nu{\rm h}{\bf k}}}+{V^{2}_{\nu{\rm h}{\bf k}}\over\omega+E_{\nu{\rm h}{\bf k}}}\right ), \label{hole-diagonal-Green's-function-1}\\
\Gamma^{\dagger}(\bf k,\omega)&=&\sum_{\nu=1,2}(-1)^{\nu}{\alpha_{\nu{\bf k}}\bar{\Delta}_{\rm h}({\bf k})\over 2 E_{\nu{\rm h}{\bf k}}}\nonumber\\
&\times&\left ({1\over\omega-E_{\nu{\rm h}{\bf k}}}-{1\over\omega+E_{\nu{\rm h}{\bf k}}}\right ),~~~~~~
\end{eqnarray}
\end{subequations}
where $\nu=1,2$, $\alpha_{\nu{\bf k}}$, $M_{\bf k}$, $\bar{\Delta}_{\rm pg}({\bf k})$, $\bar{\Delta}_{\rm h}$, the coherence factors $U_{\nu{\rm h}{\bf k}}$ and
$V_{\nu{\rm h}{\bf k}}$, and the charge carrier quasiparticle spectrum $E_{\nu{\rm h}{\bf k}}$ have been given in Ref. \onlinecite{Feng12}.

In the framework of the CSS fermion-spin theory \cite{Feng04}, the physical electron operator is given by a composite one. In this case, the d-wave charge carrier pairing state based on the exchange of the spin excitations also leads to form a d-wave electron Cooper pairing state \cite{Feng0306} due to the charge-spin recombination \cite{Anderson91}. This follows a fact that the electron Green's function is a convolution of the spin Green's function and charge carrier Green's function in the CSS fermion-spin representation \cite{Guo07}. In particular, the electron diagonal Green's function in the present case is evaluated explicitly in terms of the charge carrier diagonal Green's function (\ref{hole-diagonal-Green's-function-1}) and spin Green's function $D^{(0)-1}({\bf k},\omega)=(\omega^{2}-\omega^{2}_{\bf k})/B_{\bf k}$ as \cite{Zhao12},
\begin{widetext}
\begin{eqnarray}\label{electron-Green-function}
G({\bf k},\omega)&=&{1\over N}\sum_{{\bf p},\nu=1,2}{B_{{\bf p}+{\bf k}}\over 2\omega_{{\bf p}+{\bf k}}}\left [U^{2}_{\nu{\rm h}{\bf p}}\left ({n_{\rm F}(E_{\nu{\rm h}{\bf p}})
+n_{\rm B}(\omega_{{\bf p}+{\bf k}})\over \omega+E_{\nu{\rm h}{\bf p}}-\omega_{{\bf p}+{\bf k}}}+{1-n_{\rm F}(E_{\nu{\rm h}{\bf p}})+n_{\rm B}(\omega_{{\bf p}+{\bf k}}) \over\omega+E_{\nu{\rm h}{\bf p}}+\omega_{{\bf p}+{\bf k}}}\right )\right .\nonumber\\
&+&\left . V^{2}_{\nu{\rm h}{\bf p}}\left ({1-n_{\rm F}(E_{\nu{\rm h}{\bf p}})+n_{\rm B}(\omega_{{\bf p}+{\bf k}})\over\omega-E_{\nu{\rm h}{\bf p}}-\omega_{{\bf p}+{\bf k}}}
+{n_{\rm F}(E_{\nu{\rm h}{\bf p}})+n_{\rm B}(\omega_{{\bf p}+{\bf k}})\over\omega-E_{\nu{\rm h}{\bf p}}+\omega_{{\bf p}+{\bf k}}}\right )\right ], ~~~~~~
\end{eqnarray}
\end{widetext}
where $n_{\rm B}(\omega)$ and $n_{\rm F}(\omega)$ are the boson and fermion distribution functions, respectively, while the spin excitation spectrum $\omega_{\bf p}$ and the function $B_{\bf p}$ have been given in Ref. \onlinecite{Guo07}.

It has been shown from the angle-resolved photoemission spectroscopy (ARPES) experimental data \cite{Norman98,Kanigel06} that in the underdoped regime, although the normal-state of cuprate superconductors is metallic, the part of the Fermi surface is gapped out by the normal-state pseudogap, then the low-energy electron excitations occupy disconnected segments called as the Fermi arcs located around ${\bf k}\approx [(1-\delta)\pi/2,(1-\delta)\pi/2]$ in the Brillouin zone. In particular, in corresponding to the doping and temperature dependence of the normal-state pseudogap, the ARPES experimental results \cite{Norman98,Kanigel06,Nakayama09,Kondo09} indicate that the Fermi arc in the underdoped regime increases in length with temperatures, till at about the normal-state pseudogap crossover temperature $T^{*}$, then it covers the full length of the Fermi surface (a continuous contour in momentum space) for the temperature $T>T^{*}$. Furthermore, the Fermi arc increases its length as a function of doping \cite{Meng11}, and then it evolves into a continuous contour in momentum space near the end of the SC dome. In this case, the Fermi wave vector can be estimated qualitatively as ${\bf k}_{\rm F}\approx [(1-\delta)\pi/2,(1-\delta)\pi/2]$, and then the Fermi energy is determined \cite{Zhao12} by the pole of the electron Green's function (\ref{electron-Green-function}).

The electron spectral function $A_{\rm S}({\bf k},\omega)=-2{\rm Im}G({\bf k},\omega)$ is directly related to the analytically continued electron diagonal Green's function (\ref{electron-Green-function}) as,
\begin{widetext}
\begin{eqnarray}\label{electron-spectral-function}
A_{\rm S}({\bf k},\omega)&=&2\pi {1\over N}\sum_{{\bf p},\nu=1,2}{B_{{\bf p}+{\bf k}}\over 2\omega_{{\bf p}+{\bf k}}}\{ U^{2}_{\nu{\rm h}{\bf p}}[n_{\rm F}(E_{\nu{\rm h}{\bf p}})
+n_{\rm B}(\omega_{{\bf p}+{\bf k}})]\delta(\omega+E_{\nu{\rm h}{\bf p}}-\omega_{{\bf p}+{\bf k}})\nonumber\\
&+&U^{2}_{\nu{\rm h}{\bf p}}[1-n_{\rm F}(E_{\nu{\rm h}{\bf p}})+n_{\rm B}(\omega_{{\bf p}+{\bf k}})]\delta(\omega+E_{\nu{\rm h}{\bf p}}+\omega_{{\bf p}+{\bf k}})
+V^{2}_{\nu{\rm h}{\bf p}}[1-n_{\rm F}(E_{\nu{\rm h}{\bf p}})+n_{\rm B}(\omega_{{\bf p}+{\bf k}})]\delta(\omega-E_{\nu{\rm h}{\bf p}}-\omega_{{\bf p}+{\bf k}})\nonumber\\
&+&V^{2}_{\nu{\rm h}{\bf p}}[n_{\rm F}(E_{\nu{\rm h}{\bf p}})+n_{\rm B}(\omega_{{\bf p}+{\bf k}})]\delta(\omega-E_{\nu{\rm h}{\bf p}}+\omega_{{\bf p}+{\bf k}})\}.
\end{eqnarray}
\end{widetext}
In principle, it is straightforward to calculate the tunneling density of states once the electron spectral function throughout the Brillouin zone is known. The only complication arises from the tunneling matrix element $T_{\bf kp}$, which can be very anisotropic in cuprate superconductors. From the cuprate superconductor-insulator-normal metal (SIN) tunneling current \cite{Eschrig00},
\begin{eqnarray}\label{current}
{\rm I(V)}&=&{1\over N}\sum_{\bf k}|\bar{M}_{\bf k}|^{2}\nonumber\\
&\times&\int_{-\infty}^{\infty}{{\rm d}\omega\over 2\pi}A_{\rm S}({\bf k},\omega)[n_{\rm F}(\omega)-n_{\rm F}(\omega+e{\rm V})],
\end{eqnarray}
we can obtain the differential conductance ${\rm dI}/{\rm dV}$, where $|\bar{M}_{\bf k}|^{2}=2e(1/N)\sum_{\bf p}|T_{\bf kp}|^{2}A_{\rm N}({\bf p},\omega)$ is the SIN matrix element, while $A_{\rm N}({\bf p},\omega)$ is the spectral function of the normal metal just as the tip in the STM/STS experiments \cite{Fischer07,Pan00}. This SIN matrix element $\bar{M}_{\bf k}$ has been assumed to be energy independent, since it varies slowly with energy. For the simplicity, the tunneling matrix element $\bar{M}_{\bf k}$ has been modeled as $|\bar{M}_{\bf k}|^{2}= \bar{M}_{0}^{2}$.

\begin{figure}[h!]
\center\includegraphics[scale=0.35]{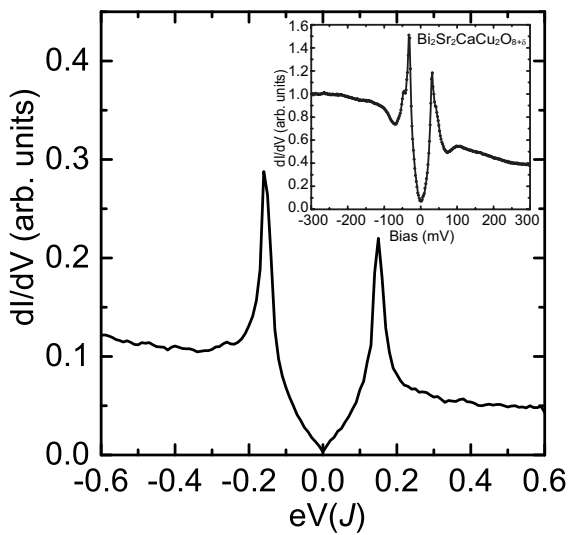}
\caption{The differential tunneling conductance spectrum at $\delta=0.09$ with $T=0.002J$ for $t/J=2.5$, $t'/t=0.3$, and $J=110$meV. Inset: the corresponding experimental data of the underdoped Bi$_{2}$Sr$_{2}$CaCu$_{2}$O$_{8+\delta}$ taken from Ref. \onlinecite{Pan00}. \label{fig1}}
\end{figure}

In Fig. \ref{fig1}, we plot the differential tunneling conductivity ${\rm dI}/{\rm dV}$ at the doping concentration $\delta=0.09$ with temperature $T=0.002J$ for parameters $t/J=2.5$, $t'/t=0.3$, and $J=110$meV. For comparison, the corresponding experimental result \cite{Pan00} of the underdoped Bi$_{2}$Sr$_{2}$CaCu$_{2}$O$_{8+\delta}$ is also plotted in Fig. \ref{fig1} (inset). It is shown clearly that our present theoretical result captures the qualitative feature of the differential tunneling spectrum observed experimentally on cuprate superconductors \cite{Fujita12,Fischer07,Renner98,Pan00,Hanaguri04,Kugler06}. In the gap region, the tunneling spectrum indicates well developed the SC coherence peaks near the gap edges $\pm\bar{\Delta}_{\rm h}$, while the finite conductance below the gap is thus consistent with the presence of the nodes in the d-wave gap. However, the most striking feature is that the tunneling conductivity is markedly asymmetric between positive and negative voltage biases, which is much different from the symmetric tunneling conductivity in the conventional superconductors \cite{Schrieffer64,Giaever60}.

\begin{figure}[h!]
\center\includegraphics[scale=0.35]{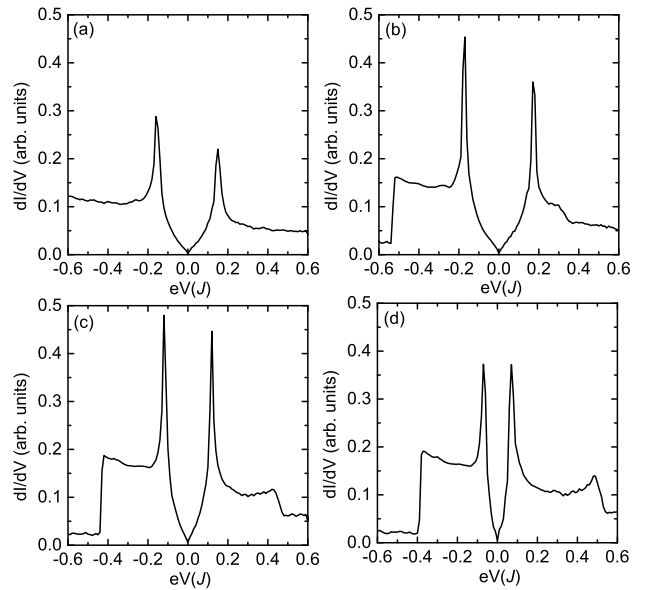}
\caption{The doping dependence of the differential tunneling conductance spectra at (a) $\delta=0.09$, (b) $\delta=0.15$, (c) $\delta=0.21$, and (d) $\delta=0.24$ with $T=0.002J$ for $t/J=2.5$, $t'/t=0.3$, and $J=110$meV. \label{fig2}}
\end{figure}

\begin{figure}[h!]
\center\includegraphics[scale=0.5]{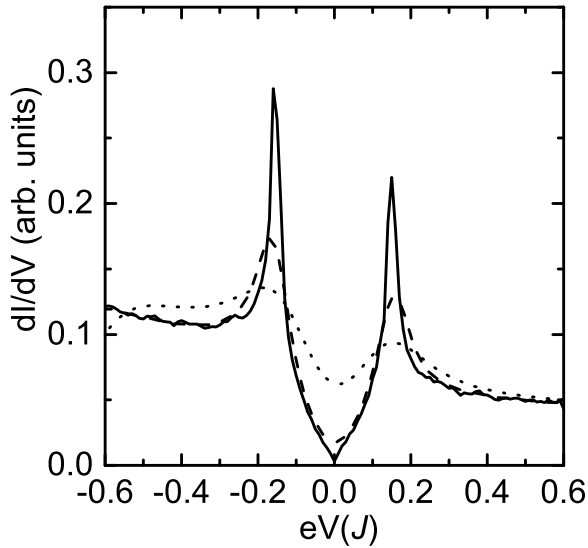}
\caption{The temperature dependence of the differential tunneling conductivity at $\delta=0.09$ with (a) $T=0.002J$ (solid line), (b) $T=0.02J$ (dashed line), and (c) $T=0.05J$ (dotted line) for $t/J=2.5$, $t'/t=0.3$, and $J=110$meV. \label{fig3}}
\end{figure}

For a better understanding of the evolution of the asymmetric tunneling spectrum with doping, we have further performed a calculation for ${\rm dI}/{\rm dV}$ at different doping concentrations, and the results of ${\rm dI}/{\rm dV}$ at (a) $\delta=0.09$, (b) $\delta=0.15$, (c) $\delta=0.21$ and (d) $\delta=0.24$ with $T=0.002J$ for $t/J=2.5$, $t'/t=0.3$, and $J=110$meV are plotted in Fig. \ref{fig2}. As a natural consequence of the domelike shape of the doping dependence of $\bar{\Delta}_{\rm h}$, the SC coherence peaks near the gap edges $\pm\bar{\Delta}_{\rm h}$ move away from the Fermi energy with increasing doping in the underdoped regime, and the distance between the SC coherence peak and Fermi energy exhibits a maximum around the optimal doping, then these SC coherence peaks are shifted toward to the Fermi energy in the overdoped regime. Moreover, although the tunneling conductivity is heavily asymmetric with respect to the Fermi energy in the underdoped and optimally doped regimes, the asymmetry of the tunneling conductivity weakens with increasing doping, and therefore there is a tendency towards to the symmetric tunneling conductivity. This tendency is particularly obvious in the overdoped regime, in particular, the asymmetric tunneling conductivity disappears in the heavily overdoped regime, and then the symmetric tunneling conductivity emergences. These results are qualitatively consistent with the experimental data observed on cuprate superconductors \cite{Fischer07,Kugler06}. Furthermore, we have discussed the temperature dependence of the tunneling conductivity in cuprate superconductors, and the results of ${\rm dI}/{\rm dV}$ at $\delta=0.09$ with (a) $T=0.002J$ (solid line), (b) $T=0.02J$ (dashed line), and (c) $T=0.05J$ (dotted line) for $t/J=2.5$, $t'/t=0.3$, and $J=110$meV are plotted in Fig. \ref{fig3}. Within the kinetic energy driven SC mechanism, the calculated \cite{Feng12} $T_{\rm c}\sim 0.053J$ for $\delta=0.09$. With increasing temperatures, the heavily asymmetric tunneling conductivity occurred in the underdoped regime at the temperature $T<<T_{\rm c}$ abates at the temperature range $T<T_{\rm c}$. In particular, the weight of the SC coherence peaks (then the asymmetric tunneling conductivity in the SC-state) follows qualitatively a charge carrier pair gap type temperature dependence, and vanishes at $T=T_{c}$, which are also in qualitative agreement with experimental data of cuprate superconductors \cite{Fischer07,Renner98}.

\begin{figure}[h!]
\center\includegraphics[scale=0.4]{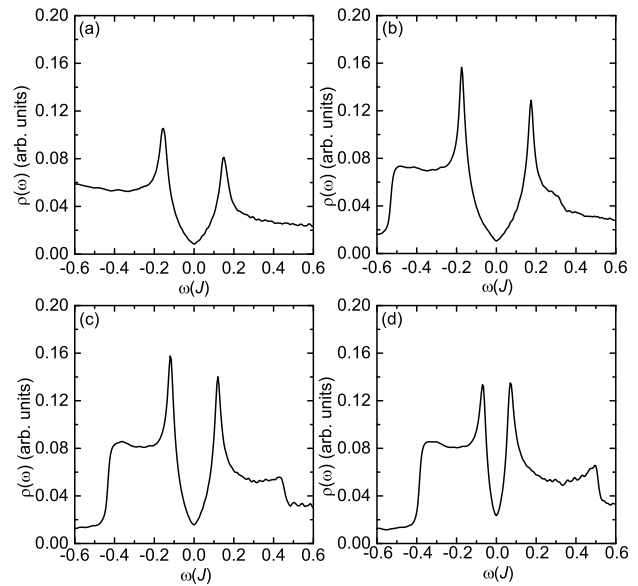}
\caption{The doping dependence of the quasiparticle density of states at (a) $\delta=0.09$, (b) $\delta=0.15$, (c) $\delta=0.21$, and (d) $\delta=0.24$ with $T=0.002J$ for $t/J=2.5$, $t'/t=0.3$, and $J=110$meV. \label{fig4}}
\end{figure}

In the conventional superconductors, the Bogoliubov quasiparticle is a coherent combination of particle (electron) and its absence (hole), i.e., its annihilation operator is a linear combination of particle and hole operators as \cite{Schrieffer64} $\beta_{\bf k}=U_{\bf k}C_{{\bf k}\uparrow}+V_{\bf k}C^{\dag}_{-{\bf k}\downarrow}$, with the constraint for the coherence factors $|U_{\bf k}|^2+|V_{\bf k} |^{2}=1$ for any wave vector ${\bf k}$ (normalization). In this case, the Bogoliubov quasiparticle do not carry definite charge. In other words, the Bogoliubov quasiparticles are not pure electrons or holes but mixtures of the two, and precisely at the gap energy they are equal mixtures, so that at the gap energy the tunneling conductivity for +${\rm V}$ and -${\rm V}$ should be equal \cite{Anderson06}. The relevant tunneling current can flow either in the form of right-moving holes or left-moving electrons, and then in the SC-state it is an equal coherent mixture of two. On the other hand, in spite of the unconventional SC mechanism, the ARPES experimental results have unambiguously established the Bogoliubov quasiparticle nature of the sharp SC quasiparticle peak in cuprate superconductors \cite{Matsui03}. However, an energy gap called the normal-state pseudogap exists \cite{Batlogg94,Timusk99,Hufner08} above $T_{\rm c}$ but below the pseudogap crossover temperature $T^{*}$. In this case, the essential physics of the asymmetry of the tunneling conductivity in cuprate superconductors is closely related to the emergence of the normal-state pseudogap \cite{Feng12}. This follows a fact that in the framework of the kinetic energy driven SC mechanism \cite{Feng0306,Feng12}, the normal-state pseudogap state is particularly obvious in the underdoped regime, i.e., the magnitude of the normal-state pseudogap is much larger than that of the pair gap in the underdoped regime, then it smoothly decreases upon increasing doping. Although the quasiparticle density of states (DOS) at both positive and negative energies are suppressed by this normal-state pseudogap, the suppression of the quasiparticle DOS at the positive energy side is more severe than the case at the negative energies. Moreover, this suppression of the quasiparticle DOS follows the same doping dependent behavior of the normal-state pseudogap, i.e., it smoothly decreases with increasing doping. To show this point clearly, we have calculated the doping dependence of the quasiparticle DOS $\rho(\omega)=(1/N)\sum_{\bf k}A_{\rm S}({\bf k},\omega)$ at different doping concentrations, and the results of $\rho(\omega)$ at (a) $\delta=0.09$, (b) $\delta=0.15$, (c) $\delta=0.21$ and (d) $\delta=0.24$ with $T=0.002J$ for $t/J=2.5$, $t'/t=0.3$, and $J=110$meV are plotted in Fig. \ref{fig4}. Our results in Fig. \ref{fig4} show clearly that the asymmetry of the quasiparticle DOS between positive and negative energies in the underdoped and optimally doped regimes weakens with increasing doping. In particular, in the heavily overdoped regime, $\bar{\Delta}_{\rm pg}\approx 0$, i.e., the effect of the normal-state pseudogap is negligible, and then the full charge carrier diagonal and off-diagonal Green's functions in Eq. (\ref{hole-Green-function-1}) can be reduced as a simple d-wave BCS formalism \cite{Feng12,Zhao12},
\begin{subequations}\label{BCSform}
\begin{eqnarray}
g({\bf k},\omega)&=&{U^{2}_{{\rm h}{\bf k}}\over\omega-E_{{\rm h}{\bf k}}}+{V^{2}_{{\rm h}{\bf k}}\over\omega+E_{{\rm h}{\bf k}}}, \\
\Gamma^{\dagger}({\bf k},\omega)&=&-{\bar{\Delta}_{\rm h}({\bf k})\over 2E_{{\rm h}{\bf k}}}\left ( {1\over\omega-E_{{\rm h}{\bf k}}}-{1\over\omega + E_{{\rm h}{\bf k}}}\right ),
\end{eqnarray}
\end{subequations}
although the pairing mechanism is driven by the kinetic energy by exchanging spin excitations, where the charge carrier quasiparticle coherence factors $U^{2}_{{\rm h}{\bf k}}= (1+\xi_{{\bf k}}/E_{{\rm h}{\bf k}})/2$ and $V^{2}_{{\rm h}{\bf k}}=(1-\xi_{{\bf k}}/E_{{\rm h}{\bf k}})/2$, and the charge carrier quasiparticle spectrum
$E_{{\rm h}{\bf k}}=\sqrt{\xi^{2}_{{\bf k}}+\mid\bar{\Delta}_{\rm h}({\bf k})\mid^{2}}$. This simple BCS formalism (\ref{BCSform}) is similar to the case in the conventional superconductors \cite{Schrieffer64}, and then the symmetric quasiparticle DOS appears in the heavily overdoped regime. Since at any point the differential tunneling conductivity is proportional to the quasiparticle DOS, the doping dependent asymmetry of the quasiparticle DOS induced by the normal-state pseudogap leads to the doping dependent asymmetry of the tunneling conductivity. In other words, the asymmetric tunneling is a natural consequence due to the presence of the normal-state pseudogap.

In conclusion, within the framework of the kinetic energy driven SC mechanism, we have discussed the asymmetric tunneling in cuprate superconductors. By considering the interplay between the SC gap and normal-state pseudogap, we have qualitatively reproduced some essential features of the doping and temperature dependence of the asymmetric tunneling. In particular, the asymmetry of the tunneling spectrum in the underdoped regime weakens with increasing doping, and then the symmetric tunneling spectrum recovers in the heavily overdoped regime. Our results also show that the asymmetric tunneling is a natural consequence due to the presence of the normal-state pseudogap.

\acknowledgments

The authors would like to thank Professor Zhi Wang for helpful discussions. This work was supported by the funds from the Ministry of Science and Technology of China under Grant Nos. 2011CB921700 and 2012CB821403, and the National Natural Science Foundation of China under Grant No. 11274044.

\end{document}